\input harvmac
\def\ev#1{\langle#1\rangle}
\def\np#1#2#3{Nucl. Phys. B{#1} (#2) #3}
\def\pl#1#2#3{Phys. Lett. B{#1} (#2) #3}
\def\plb#1#2#3{Phys. Lett. B{#1} (#2) #3}
\def\prl#1#2#3{Phys. Rev. Lett. {#1} (#2) #3}
\def\physrev#1#2#3{Phys. Rev. D{#1} (#2) #3}

\def\f#1#2{\textstyle{#1\over #2}}
\def\drawbox#1#2{\hrule height#2pt 
        \hbox{\vrule width#2pt height#1pt \kern#1pt 
              \vrule width#2pt}
              \hrule height#2pt}

\def\Asym#1#2{\vcenter{\vbox{\drawbox{#1}{#2}
              \kern-#2pt       
              \drawbox{#1}{#2}}}}

\batchmode
  \font\bbbfont=msbm10
\errorstopmode
\newif\ifamsf\amsftrue
\ifx\bbbfont\nullfont
  \amsffalse
\fi
\ifamsf
\def\IR{\hbox{\bbbfont R}} 
\def\IC{\hbox{\bbbfont C}}

\def\IZ{\hbox{\bbbfont Z}}
\def\IF{\hbox{\bbbfont F}}
\def\IP{\hbox{\bbbfont P}}
\else
\def\IR{\relax{\rm I\kern-.18em R}}
\def\IC{\relax{\rm {\bf C}}}
\def\IZ{\relax\ifmmode\hbox{Z\kern-.4em Z}\else{Z\kern-.4em Z}\fi}
\def\IF{\relax{\rm I\kern-.18em F}}
\def\IP{\relax{\rm I\kern-.18em P}}
\fi
\def\N#1#2{${\cal N}=(#1, #2)$}

\def\FI{Fayet-Iliopoulos}
\def\lfm#1{\medskip\noindent\item{#1}}
\lref\Wi{E. Witten, hep-th/9503124, \np{443}{1995}{85}.}
\lref\Wii{E. Witten, hep-th/9511030, \np{460}{1996}{541}.}
\lref\snewt{N. Seiberg, hep-th/9705221, Phys. Lett B408 (1997) 98.}
\lref\inewt{K. Intriligator, hep-th/9708117, Adv. Theor. Math. Phys. 1
(1997) 271.}
\lref\OGSS{O.J. Ganor and S. Sethi, hep-th/9712071, J. High. Energy
Phys. 01 (1998) 007.}
\lref\NStd{N. Seiberg, hep-th/9606017, \plb{384}{1996}{81}.}
\lref\ismir{K. Intriligator, N. Seiberg, hep-th/9607207,
\plb{387}{1996}{513}.} 
\lref\berk{J. de Boer, K. Hori, H. Ooguri, Y. Oz, hep-th/9611063,
\np{493}{1997}{101}.}
\lref\sen{A. Sen, hep-th/9605150, \np{475}{1996}{562}.}
\lref\BDS{T. Banks, M. Douglas, and N. Seiberg, hep-th/9605199,
\plb{387}{1996}{278}.}
\lref\mrdi{M.R. Douglas, hep-th/9512077.}
\lref\SW{N. Seiberg and E. Witten, hep-th/9607163.}
\lref\prince{O.J. Ganor two papers}
\lref\wsem{E. Witten, seminar at Aspen Center for Physics, Aug. '97.}
\lref\wgauge{E. Witten, hep-th/9710065, 
Adv. Theor. Math. Phys. 2 (1998) 61.}
\lref\MPAZ{M. Porrati and A. Zaffaroni, hep-th/9611201,
\np{490}{1997}{107}.}
\lref\BFSS{T. Banks, W. Fischler, S.H. Shenker, L. Susskind,
hep-th/9610043, \physrev{55}{1997}{5112}.}
\lref\JMAS{J. Maldacena and A. Strominger, hep-th/9710014, JHEP 9712
(1997) 008.}
\lref\APJP{A. Peet and J. Polchinski, hep-th/9809022,
\physrev{59}{1999}{065011}.}
\lref\SMNS{S. Minwalla and N. Seiberg, hep-th/9904142, JHEP 9906
(1999) 007.}
\lref\PG{P. Ginsparg, \physrev{35}{1987}{648}.}
\lref\PARP{P.S. Aspinwall and M.R. Plesser, hep-th/9905036, JHEP 9908
(1999) 001.}
\lref\CGK{Y.K. Cheung, O.J. Ganor, M. Krogh, hep-th/9805045,
\np{536}{1998}{175}.}
\lref\CGKM{Y.E. Cheung, O.J. Ganor, M. Krogh, A.Y. Mikhailov,
hep-th/9812172}
\lref\EPGR{E. Perevalov and G. Rajesh, hep-th/9706005,
\prl{79}{1997}{2931}.} 
\lref\AGMPER{A. Giveon, M. Porrati and E. Rabinovici, hep-th/9401139,
Phys. Rept. 244 (1994) 77.}
\lref\vafa{C. Vafa, hep-th/9602022, \np{469}{1996}{403}.}
\lref\SWii{N. Seiberg and E. Witten, hep-th/9408009,
\np{431}{1994}{484}.}
\lref\wcomm{E. Witten, hep-th/9507121, Proc. Strings '95.}
\lref\BRSI{M. Berkooz, M. Rozali, and N. Seiberg, hep-th/9704089,
\pl{408}{1997}{105}.}
\lref\PADMO{P.S. Aspinwall and D.R. Morrison, hep-th/9404151.}
\Title{hep-th/9909219,  UCSD/PTH 99-12, IASSNS-HEP-99/86}
{\vbox{\centerline{Compactified Little String Theories and}
\centerline{Compact Moduli Spaces of Vacua
}}}
\medskip
\centerline{Kenneth Intriligator}
\vglue .5cm
\centerline{UCSD Physics Department}
\centerline{9500 Gilman Drive}
\centerline{La Jolla, CA 92093}
\vglue .25cm
\centerline{and}
\vglue .25cm
\centerline{School of Natural Sciences\footnote{${}^*$}{address for
Fall term, 1999.}}
\centerline{Institute for Advanced Study}
\centerline{Princeton, NJ 08540, USA}

\bigskip
\noindent

It is emphasized that compactified little string theories have compact
moduli spaces of vacua, which globally probe compact string
geometry. Compactifying various little string theories on $T^3$ leads
to 3d theories with exact, quantum Coulomb branch given by: an
arbitrary $T^4$ of volume $M_s^2$, an arbitrary K3 of volume $M_s^2$,
and moduli spaces of $G=SU(N)$, $SO(2N)$, or $E_{6,7,8}$ instantons on
an arbitrary $T^4$ or K3 of fixed volume.  Compactifying instead on a
$T^2$ leads to 4d theories with a compact Coulomb branch base which,
when combined with the exact photon gauge coupling fiber, is a compact,
elliptically-fibered space related to the above spaces.

\Date{9/99}

\newsec{Introduction}

Over the past few years, there have been a variety of connections
between the moduli spaces of supersymmetric gauge theories and stringy
geometry.  For example, singular background geometry or gauge bundles
can lead to enhanced, non-perturbative, gauge theories, whose moduli
spaces reproduce the local singularity \refs{\Wi, \Wii}.  Another
connection is via branes, whose world-volume supersymmetric gauge
theory has a moduli space which ``probes'' \refs{\mrdi, \BDS} the
geometry in which the branes live.  In an extreme form of this
connection \BFSS, we perhaps actually live in the moduli space of a
supersymmetric theory.  It is thus interesting to consider, generally,
what types of geometry can be reproduced via moduli spaces of vacua.

A basic issue is whether moduli spaces of vacua can be compact.
Moduli spaces of vacua of standard gauge theories are generally
non-compact cones: if a given set of scalar expectation values
$\ev{\phi _i}$ is a D-flat vacuum, so is $\lambda
\ev{\phi _i}$ for arbitrary scaling factor $\lambda \rightarrow
\infty$.  (This is slightly modified by \FI\ terms for
$U(1)$ factors.)  An exception is the Coulomb branch moduli,
associated with the Wilson lines, of gauge theories which are
compactified on tori; these moduli live on dual tori, modded out by
the Weyl group.

The present note is devoted to emphasizing that toroidally
compactified ``little string theories'' \refs{\BRSI, \snewt} \foot{The
extent to which these 6d theories decouple from the 10d bulk, for
energies above some gap value, is subtle \refs{\JMAS, \APJP, \SMNS};
we will ignore these issues and only discuss the vacuum manifold.}
can have a variety of interesting, compact, moduli spaces of vacua.
The present discussion is an elaboration of a footnote which appeared
in
\inewt.  The basic message is that, while world-volume gauge
theories only {\it locally} probe the geometry transverse to the
brane, little string extensions can {\it globally} probe {\it compact}
geometry.  While this fact is perhaps well-known to some experts, it
is hoped that some readers will find it of interest.  

For example, the basic ${\cal N}=(1,0)$ heterotic little string
theory, when compactified on $T^3$, is argued to have a Coulomb branch
moduli space of vacua which is a K3 of volume $M_s^2$.  (Since a 3d
scalar has mass dimension $\half$, this has the correct dimensions.)
The Coulomb branch is a non-linear sigma model with exact, quantum
metric equal to the Ricci-flat metric of K3.  The K3 is an arbitrary
one of fixed volume, whose parameter space coincides with that of the
$T^3$ compactified heterotic little string theory; the map between
these parameter spaces is the same as enters in the duality between
the 10d heterotic string on $T^3$ and M theory on K3.  Geometric
symmetries of the K3 Coulomb branch map to non-trivial T-dualities of
the $T^3$ compactified little string theory.

More generally, it will be argued that the little string theories
obtained in \inewt\ from $K$ heterotic (or type II) NS branes at a
transverse $\IC ^2/\Gamma _G$ singularity, when compactified on $T^3$,
have a compact Coulomb branch moduli space of vacua given by the
moduli space of $K$ $G$-instantons on K3 (or $T^4$).  Here $G$ is an
arbitrary $A,D,E$ group and $\Gamma _G$ is the corresponding $SU(2)$
subgroup.  The K3 or $T^4$ appearing here is precisely that of $M$
theory duality, which the compactified little string theory globally
probes.  The volume of the compact Coulomb branch is again set by
$M_s$.  In each case, the Coulomb branch sigma model metric must be
the unique one which is Ricci flat.

Similarly, it will be argued that little string theories, when
compactified to 4d on a $T^2$, have Coulomb branches which globally
probe $F$ theory.  The Coulomb branch is the base space of $F$ theory,
and the photon kinetic terms are the elliptic fibration.  For example,
compactifying the basic ${\cal N}=(1,0)$ little string theory on $T^2$
leads to a 4d theory whose total space of Coulomb branch base and
Seiberg-Witten curve is an elliptically fibered K3 of volume $M_s^2$.
The map between the $T^2$ compactification data and the parameter
space of fixed volume, elliptically fibered, K3 spaces is the same as
in the duality between the 10d heterotic theory on $T^2$ and $F$
theory on an elliptically fibered K3 \vafa.

The next section will review little string theories and their
compactification, with several new minor comments included.  Sect. 3
outlines classical $T^3$ dimensional reduction of ordinary
6d $U(1)$ and $SU(2)$ gauge fields.  This already leads to
compact Coulomb branches, of fixed volume $g_6^{-2}$; the Coulomb
branch for $U(1)$ is $T^4$, while that of $SU(2)$ is K3. Sect. 4
extends the probe argument of \NStd\ to argue for our main message:
that compactifying little string theories on $T^3$ leads to theories
whose exact Coulomb branch is compact and globally probes the $T^4$ or
$K3$ of M theory duality.  Sect. 4 discusses $T^2$ compactification of
little string theories, which similarly have compact Coulomb branches
that globally probe compactification of $F$ theory, e.g. on
elliptically fibered K3s.

The proposed relation of \inewt\ between compactified type II little
string theories and moduli spaces of instantons on $T^4$ also entered
in \refs{\OGSS, \CGK, \CGKM}, where it was extended to moduli spaces
of instantons on a non-commutative $T^4$ by introducing R-symmetry
twists in the compactification.  A relation between the twisted,
compactified $(2,0)$ theory and K3 was proposed in
\CGK; this appears to be unrelated to the presently discussed
appearance of K3 in the context of the untwisted, compactified,
heterotic little string theories.  Much as in \refs{\CGK, \CGKM}, it
should also be possible to introduce R-symmetry twists for the
compactified heterotic little string theories, perhaps leading to
moduli spaces of instantons on a non-commutative K3, though this will
not be done here.

\newsec{Review of little string theories and their compactification}

Four classes of 6d little string theories were obtained in \snewt\ via
the world-volume of 5-branes in the limit $g_s\rightarrow 0$ with
$M_s$ held fixed:

\lfm{(iia)} \N{1}{1}\ supersymmetric, via $K$ IIB NS five-branes
\snewt\ or via IIA or M theory with a $\IC ^2/\Gamma _G$ ALE 
singularity \refs{\wsem, \wgauge}.

\lfm{(iib)} \N{2}{0}\ supersymmetric, 
via $K$ type IIA five branes \snewt\ or type IIB with a $\IC ^2/\Gamma
_G$ singularity \wsem.

\lfm{(o)} \N{1}{0}\ supersymmetric, via $K$ $SO(32)$ heterotic
small instantons.  

\lfm{(e)} \N{1}{0}\ supersymmetric, via $K$ heterotic 
$E_8\times E_8$ small instantons.

Cases (iia) and (o) contain gauge fields with coupling
$g_6^{-2}=M_s^2$ and are IR free.  Instantons in the 6d gauge theories
are fundamental strings, with tension $g_6^{-2}=M_s^2$.  Cases (iib)
and (e) instead contain tensor multiplet two-form gauge fields, with
self-dual field strength, and lead to interacting RG fixed point field
theories in the IR.

${\cal N}=(1,0)$ tensor multiplet theories (of which ${\cal N}=(2,0)$
is a special case) always have an associated group $G$.  For cases
(iib) it is $SU(K)$ or the $ADE$ singularity group $G$, while for (e)
it is $Sp(K)$.  There is a $r=$rank$(G)$ dimensional, compact Coulomb
branch moduli space, with the real scalars in the ${\cal N}=(1,0)$
tensor multiplets taking values $\ev{\vec \Phi}$ in the ``$G$-Coxeter
box'' $(S^1)^{\otimes r}/W_G$, where the $S_1$ is of radius $M_s^2$
and $W_G$ is the Weyl group of $G$.  The theory is interacting at the
boundaries of the Coxeter box but, in the bulk, behaves in the IR as
$r$ free self-dual tensor multiplets.  Strings are charged under the
$r$ 2-form gauge fields of these tensor multiplets, with charge
vectors $\vec \alpha$ in the $G$ root lattice\foot{Because of the
self-duality, these strings can be regarded as either ``electrically''
or ``magnetically'' charged.  The Dirac quantization condition thus
implies that the lattice $\Lambda$ must be an {\it integer lattice},
i.e. the dot product of any two lattice vectors is an integer, so
$\Lambda \subset \widetilde
\Lambda$, where $\widetilde \Lambda$ is the dual lattice.  This is, of
course, a weaker condition than self-duality of the lattice.  For
example, the root lattice of a simple group $G$ is generally not
self-dual but, rather, a subgroup, of degree given by the center of
$G$, in the dual lattice, which is the weight lattice.}. Via a BPS
formula, a string with charges $\vec \alpha$ has tension $Z=\vec
\alpha \cdot \vec \Phi$, becoming tensionless at the origin of the
Coulomb branch.  Reducing to 5d leads to a gauge theory with
non-Abelian gauge group $G$ at the origin, so the 6d theory can be
regarded as a non-Abelian self-dual two-form gauge theory with group
$G$ (whatever that means).

\nref\PSA{P.S. Aspinwall, hep-th/9612108, \np{496}{1997}{149}.}
\nref\obrane{K. Intriligator, hep-th/9702038, \np{496}{1997}{177}.}
\nref\obraneii{J.D. Blum and K. Intriligator, hep-th/9705030,
\np{506}{1997}{223}; J.D. Blum and K. Intriligator, hep-th/9705044,
\np{506}{1997}{199}.}
\nref\PADM{P.S. Aspinwall and D.R. Morrison, hep-th/9705104, 
\np{503}{1997}{533}.}

Each of the four above classes has either vector or tensor multiplets,
but not both.  Theories containing both vector and tensor multiplets
were discussed in \inewt\ by combining 5-branes with $\IC ^2/\Gamma _G$
orbifold singularities in the transverse dimensions, using results
obtained in \refs{\PSA - \PADM}.  In this way, new
theories can be obtained for each of the four classes of branes, type
IIA, IIB, $SO(32)$ heterotic, and $E_8\times E_8$ heterotic, at
$\IC ^2/\Gamma _G$ singularities.  All of these theories generally have
${\cal N}=(1,0)$ supersymmetry.

E.g. $K$ type IIB NS 5-branes at a $\IC ^2/\Gamma _G$ singularity
\obraneii\ has
a quiver gauge theory, based on the extended Dynkin diagram of the
$ADE$ singularity group $G$, with gauge group $U(1)_D\times \prod
_{\mu =0}^r SU(Kn_\mu )$ and bi-fundamental matter. $n_\mu$ are the
$G$ Dynkin indices and $r=$rank$(G)$.  There are $r$ ${\cal N}=(1,0)$
tensor multiplets, which are associated, as described above, with the
singularity group $G$. Via an anomaly cancellation mechanism,
$SU(Kn_\mu)$ has gauge coupling $g_{\mu ,eff}^{-2}= M_s^2\delta _{\mu
, 0}+\vec \alpha _\mu \cdot \vec \Phi$ and an $SU(Kn_\mu )$ instanton,
which is a string in 6d, has tensor-multiplet charges $\vec \alpha
_\mu$ and BPS tension $Z_\mu = g_{\mu ,eff}^{-2}$.  Here the $\vec
\alpha _\mu$ are the $G$ root vectors ($\vec \alpha _0$ is the
extending root) and the condition that all $g_{\mu , eff}^{-2}\geq 0$
is precisely that the Coulomb branch $\ev{\vec \Phi}$ is the $G$
Coxeter box, of side length $M_s^2$.  The $SU(Kn_\mu )$ instanton
string charges span the $G$ root lattice.

The instanton string for a diagonal $SU(K)_D\subset
\prod _{\mu =0}^r SU(Kn_\mu)$, with index of embedding $n_\mu$ in
$SU(Kn_\mu)$, has tension $n^\mu Z_\mu=M_s^2$, and is identified with
the fundamental IIB string.  The other $r$ independent instanton
strings in $\prod _{\mu =0}^r SU(Kn_\mu )$ are to be identified with
the strings obtained \wcomm\ by wrapping the type IIB 3-brane on the
$r$ independent, fully collapsed, two-cycles of the $\IC ^2/\Gamma _G$
singularity; $m=1\dots r$ of these strings become tensionless for
$\ev{\vec \Phi}$ at a codimension $m$ boundary of the Coulomb branch
Coxeter box.

The simplest heterotic case is $K$ $SO(32)$ 5-branes at a $\IC
^2/\Gamma _G$ singularity \refs{\PSA -\PADM}.  The theories are
associated \inewt\ with a subgroup $H$ of the singularity group $G$,
with $G\rightarrow H$ as $SU(2P)\rightarrow Sp(P)$,
$SO(4P+2)\rightarrow SO(4P+1)$, $SO(4P)\rightarrow SO(4P)$,
$E_6\rightarrow F_4$, $E_7\rightarrow E_7$, $E_8\rightarrow E_8$.  The
gauge group and matter content is given by a quiver diagram, which is
the extended $H$ Dynkin diagram, with $SO$, $Sp$, and $SU$ groups at
various nodes, e.g. the group at the $\mu =0$ node is $Sp(K)$.  There
are $r=$rank$(H)$ ${\cal N}=(1,0)$ tensor multiplets, which are
associated with the group $H$.  Via an anomaly cancellation mechanism,
the gauge group at node $\mu =0 \dots r$ of the quiver diagram has
coupling $g^{-2}_{\mu ,eff}=M_s^2\delta _{\mu ,0}+\vec \alpha _\mu
\cdot \vec \Phi$, and an instantons string in this group has tensor
multiplet charges $\vec
\alpha _\mu$ and BPS tension $Z_\mu =g_{\mu, eff}^{-2}$.  
Here $\vec \alpha _\mu$ are the simple and extending roots of $H$, so
the instanton strings span the $H$ root lattice.  Instantons in a
diagonal $Sp(K)_D$ are identified with the fundamental heterotic
string, of tension $M_s^2$.  The other $r$ independent instanton
strings can again be identified with 3-branes wrapped on collapsed
two-cycles; $m=1\dots r$ of these become massless at a codimension $m$
boundary of the Coulomb branch (the $H$ Coxeter box).

The other heterotic case, $K$ $E_8\times E_8$ 5-branes at a $\IC
^2/\Gamma _G$ singularity, leads to little string theories with a more
involved spectrum of tensor multiplets, gauge groups, and matter
content \refs{\PADM, \inewt}.

Compactifying on a circle, 6d vector and tensor multiplets both lead
to 5d vector multiplets.  A 6d ${\cal N}=(1,0)$ theory with a gauge
group of rank $r_V$ and $n_T$ tensor multiplets, when compactified,
leads to a 5d theory with a Coulomb branch moduli space of vacua of
dimension $d_C=r_V+n_T$.  Compactifying to 4d, the Coulomb branch has
real dimension $2(r_V+n_T)$ and in 3d, upon dualizing the $d_C$
photons, there is a Coulomb branch of real dimension $4(r_V+n_T)$.

Little string theories exhibit T-duality when compactified on a circle
\snewt, with the (iia) theory on a circle of radius $R$ identical to
the (iib) theory on a circle of radius $1/M_s^2R$.  Similarly, the (o)
heterotic theory, on a circle of radius $R$, and with a Wilson line
around the circle breaking $SO(32)$ to $SO(16)\times SO(16)$, is
identical to the (e) heterotic theory on a circle of radius
$1/M_s^2R$, again with a Wilson line breaking $E_8\times E_8$ to
$SO(16)\times SO(16)$.  (See \PG\ for the heterotic T-duality with
general Wilson lines.)  In these cases, T-duality exchanges 6d tensor
and vector multiplets, $r_V\leftrightarrow n_T$.  This is nice because
the 5d classical kinetic terms for the scalars coming from 6d tensor
multiplets, $M_s^4R(d\Phi )^2$, is indeed exchanged with the kinetic
term, $g_6^{-2}R(R^{-1}d\Phi )^2$, of a vector multiplet on a circle
of radius $R$.  In both cases $\Phi$ is a compact scalar, normalized
so $\Phi \in [0,1]$, and the two kinetic terms are exchanged by
$R\leftrightarrow (M_s^2R)^{-1}$ upon setting $g_6^{-2}=M_s^2$.  

More generally, there is an expected T-duality, with $R\leftrightarrow
(M_s^2R)^{-1}$ exchanging the theories coming from IIA and IIB or
$SO(32)$ and $E_8\times E_8$ heterotic branes at $\IC ^2/\Gamma _G$
singularities.  T-dual theories must have $r_V+n_T=\widetilde
r_V+\widetilde n_T$.  As was noted in \refs{\EPGR, \inewt}, this is the
case for the $SO(32)$ and $E_8\times E_8$ branes at singularities:
both cases have $r_V+n_T=C_2(G)K-|G|$, where $C_2(G)$ is the dual
Coxeter number of the singularity group $G$ and $|G|$ is its
dimension.  This formula will be important in what follows.  A point
of concern mentioned in \inewt\ is that a stronger condition,
$r_V=\widetilde n_T$ and $n_T=\widetilde r_V$, needed for T-duality to
exchange the classical kinetic terms as above, is not satisfied.  The
present situation is, in fact, closely connected to that of \PARP,
where it was argued that T-duality can fail.  Here, however, there is
a simpler resolution: the Coulomb branch metric can get quantum
corrections and, while the quantum corrected metrics are expected to
agree, the classical metrics need not.  The stronger condition is thus
unnecessary.

All little string theories, when compactified on $T^D$, have the
parameter space \snewt\
\eqn\parmg{O(D+y,D;\IZ)\backslash O(D+y,D)/O(D+y)\times O(D),}
where $y=0$ for the type II cases and $y=16$ for the heterotic cases.
These are the $T^D$ metric and $B_{NS}$ fields ($D^2$ real
parameters), and also the $SO(32)$ or $E_8\times E_8$ Wilson lines in
the heterotic cases ($16D$ real parameters). $O(D+y,D;\IZ)$ is the full
$T$ duality group.

\newsec{Compactification preliminaries}

We first consider the classical dimensional reduction of a
$6d$ $U(1)$ gauge field, 
\eqn\ssd{\int d^6x (-{1\over 4g_6^2} F_{\mu \nu}F^{\mu \nu}
+B_{NS}\wedge F\wedge F),} on a $T^3$ to three dimensions.  $B_{NS}$
is an external, background, two-form gauge field.  We take space to be
$\IR ^3\times T^3$, with $\IR ^3$ coordinates $x^i$, $i=1,2,3$, and
periodic coordinates $\rho ^a\in [0,1]$, $a=1,2,3$, for the $T^3$; the
metric is $ds^2=\delta _{ij}dx^idx^j+ h_{ab}d\rho ^a d\rho ^b$.
Taking all fields to be independent of the $T^3$ coordinates $\rho
^a$, \ssd\ becomes
\eqn\ssdi{S=\int d^3x \left[{\sqrt{\det h}\over g_6^2}
(-\f{1}{4}F_{ij}F^{ij}+\half (h^{-1})^{ab}\partial _i\phi _a\partial ^i
\phi _b)+\theta ^a\epsilon ^{ijk}F_{ij}\partial _k\phi _a\right],}
where $B_{NS}=\epsilon _{abc}\theta ^ad\rho ^b\wedge d\rho ^c$ for
some constants $\theta ^a$, $a=1,2,3$.  The three real scalars $\phi
_a$ are associated with the Wilson lines of the gauge field around the
cycles $d\rho ^a$ of the $T^3$ and are periodic, normalized so that
$\phi _a\in [0,1]$.
  
The 3d $U(1)$ gauge field can be dualized to another real scalar,
which also lives on a circle.  This is done as in
\SW: we replace $F_{ij}\rightarrow F_{ij}-H_{ij}$ in \ssdi\ and
introduce an additional term $\epsilon ^{ijk}H_{ij}\partial _k \phi
_4$, with the scalar $\phi _4$ periodic, normalized so that $\phi
_4\in [0,1]$.  First integrating out $\phi _4$ leads back to the
original theory.  First integrating out $H$ sets $F_{ij}=0$ and leads
to $\phi _4$ kinetic terms.  Combining with the $\phi _a$ kinetic
terms in \ssdi, the upshot is a $T^4$ Coulomb branch moduli space of
vacua $\ev{\phi _A}$, $A=1, \dots 4$, with metric
\eqn\metricf{ds^2 =
{\sqrt{\det h}\over g_6^2}(h^{-1})^{ab}d\phi _a d\phi _b +{g_6^2\over
\sqrt{\det h}}(d\phi _4-\theta ^a d\phi _a)^2\equiv G^{AB}d\phi
_A d\phi _B,} where $a$ runs over $1,2,3$ and $A=1, \dots 4$.  This
metric $G^{AB}$ has 10 real components, which depend on the 9 real
parameters $h^{ab}$ and $\theta ^a$, and thus satisfies one
constraint.  The relation is that the $T^4$ of \metricf\ has fixed
volume, independent of the $h_{ab}$ and the $\theta ^a$:
\eqn\volti{{\rm Volume}(T^4)=\sqrt{\det (G^{AB})}={1\over g_6^2}=M_s^2.}

Although the above discussion was purely classical, the map
\metricf\ between the $T^3$ metric $h_{ab}$ and $B$ fields $\theta ^a$
is exactly the relevant one for relating type IIB string theory on
$T^3$ to M theory on $T^4$, to be discussed in the next section.
Indeed, the map \metricf\ was also obtained in \CGKM\ in the context
of the compactified $(2,0)$ theory via a chain of string duality
gymnastics. 

We pause to note that the metric \metricf\ nicely exhibits properties
to be expected based on its connection to M theory.  In particular,
the obvious, geometric $SL(4;\IZ )$ discrete symmetries of the $T^4$
correspond to non-trivial T-dualities, in a subgroup of the T-duality
group appearing in \parmg.  For example, consider the obvious
requirement that the $T^4$ be invariant under the relabeling exchange
$\phi _3\leftrightarrow - \phi _4$.  Taking, for simplicity, $T^3$
with $h_{ab}=L_a^2\delta _{ab}$ and $\theta _a=0$, it follows from
\metricf\ that this operation corresponds to the operation
\eqn\texamp{L_1\rightarrow (M_s^2L_2)^{-1},\quad  L_2\rightarrow
(M_s^2 L_1)^{-1}, \quad L_3\rightarrow L_3,} where we set
$g_6^{-2}=M_s^2$.  This is a $T$ duality in two circles, which is
non-trivial but, nevertheless, a symmetry taking the IIA or IIB theory
back to itself.  The generalization of the T-duality \texamp\ for
general $h_{ab}$ and $\theta _a$ is quite complicated, see
e.g. \AGMPER; remarkably, it is indeed reproduced from \metricf\ by
simply requiring the $\phi _3\leftrightarrow - \phi _4$ symmetry.

On the other hand, T-duality in an odd number of cycles, such as the
$O(3,3;\IZ )$ element taking all $L_i\rightarrow (M_s^2L_i)^{-1}$,
for $i=1,2,3$, is not a geometric $SL(4; \IZ)$ symmetry of \metricf.
This is sensible, since such operations are not symmetries of IIA or
IIB string compactifications but, rather, exchange IIA and IIB.

In particular, starting instead from a 6d tensor multiplet,
dimensional reduction on a $T^3$ leads to a $T^4$ Coulomb branch
moduli space, with metric related to \metricf\ by T-duality in an odd
number of the $T^3$ cycles, corresponding to the exchange of IIA and
IIB.

Now consider $T^3$ reduction of a 6d $SU(2)$ gauge theory.  The above
discussion for $U(1)$ carries over to this case with almost no
changes.  The only difference is that the real scalars $\phi _A$ must
be modded out by the Weyl group action $\phi _A\sim -\phi _A$.
Modding out the $T^4$ by this $\IZ _2$ action leads to a $K3$.  Thus the
Coulomb branch of a 6d $SU(2)$ gauge theory reduced to 3d on a $T^3$
is given by $\ev{\phi _A}$ in a compact K3.  The volume of the K3 is
again set by $g_6^{-2}$, and equal to $M_s^2$.  The full parameter
space of K3 metrics of fixed volume is 57 dimensional and given by
\parmg\ with $D=3$ and $y=16$, while that obtained here only depends
on the 9 dimensional subspace given by
\parmg\ with $D=3$ and $y=0$. The remaining parameters will come from 
3 real masses for each of 16 $SU(2)$ fundamental matter flavors; these
enter as the Wilson loop parameters in \parmg.

\newsec{The probe argument, checks, and comments}

The parameter space \parmg\ for $T^3$ compactified heterotic (or type
II) little string theories coincides with the geometric parameter
space of a K3 (or $T^4$) of fixed volume.  These are, of course, the
standard miracles which enter in the duality between the 10d heterotic
(or type II) string on $T^3$ and M theory on $Y=$K3 (or $T^4$).  The
fundamental string arises as the M5 brane wrapped on $Y$, so
$M_p^6Vol(Y)=M_s^2$.  We will here extend the probe argument of
\NStd\ to argue that these M theory dualities provide the solution for
the exact, quantum, Coulomb branch metric of the $T^3$ compactified
little string theories.

Recall that the argument of \NStd\ started with 3d ${\cal N}=4$
supersymmetric (8 supercharges) $SU(2)$ gauge theory with fundamental
matter, which is the world-volume field theory in a D2 brane in type
I' string theory on $T^3$.  This maps to a M2 brane in M theory on
K3, which can be at an arbitrary point in the transverse $\IR ^4\times
K3$.  The $\IR ^4$ corresponds to a decoupled hypermultiplet in the
world-volume theory.  The $K3$ factor is more interesting: it was thus
argued in
\NStd\ that the full, quantum-corrected metric on the Coulomb branch
of the D2 brane world-volume field theory must be a local piece of the
corresponding K3; this was confirmed in \SW\ purely in the context of
3d field theory.

The D2 brane world-volume field theory only {\it locally} probes the
K3 because of the particular limit taken to decouple the bulk
dynamics: $g_s\rightarrow 0$ and $M_s\rightarrow \infty$.  On the
other hand, we can take $g_s\rightarrow 0$, but with $M_s$ held fixed.
This theory is precisely the 6d heterotic little string theory (o),
compactified to 3d on the same $T^3$ as the 10d heterotic or type I'
bulk theory.  The $T^3$ compactified little string theory (o) globally
probes the fixed volume K3 of M theory, and must thus have a Coulomb
branch moduli space of vacua which is the same K3.  The geometric K3
has volume $M_s^2M_p^{-6}$ and, taking into account how the properly
normalized Coulomb moduli scalars probe geometry, the volume of the
Coulomb branch K3 is $M_s^2$.  This matches with the result of the
previous section.  This compact Coulomb branch properly becomes
non-compact in the field theory limit $M_s^2\rightarrow \infty$.

The K3 Coulomb branch can have singularities, depending on the choice
of parameters in \parmg.  As in \NStd, these singularities mark the 
intersection of the Coulomb branch with a Higgs branch, with an
interacting 3d infra-red conformal field theory at the intersection.

Unfortunately, both sides in the present equivalence, between the
quantum Coulomb branch of the $T^3$ compactified little string theory
on the one hand, and the metric of K3 on the other, are presently not
well understood.  Perhaps the present equivalence will eventually 
be useful for using one of the two sides to learn about the other.

A direct generalization of the above is to consider a $T^3$
compactification of the little string theory (o) associated with $K$
$SO(32)$ heterotic small instanton.  This maps to $K$ M2 branes at
points on $\IR ^4\times K3$.  The Coulomb branch is, correspondingly,
the symmetric product $(K3)^{\otimes K}/S_K$, where each K3 is again
of fixed volume $M_s^2$.

The geometric symmetries (see, e.g. \PADMO) of the Coulomb branch K3
correspond to non-trivial T-dualities in \parmg\ though, as in
\texamp, only the subgroup which takes the 
$SO(32)$ heterotic theory back to itself.  An additional $\IZ _2$
component of T-dualities in $O(19,3;\IZ)$ reflects the fact that,
instead compactifying the $E_3\times E_8$ heterotic little string (e),
with T-dual $T^3$ compactification data, also yields the {\it same} 3d
theory, with the {\it same} K3 compact Coulomb branch as described
above.

Each of the little string theories reviewed in sect. 2 can be
compactified to 3d on a $T^3$, and each has an exact quantum Coulomb
branch which globally probes the dual M theory compactification.  In
each case, there is a compact Coulomb branch component, with unit
volume in units of $M_s$.  The 3d field theory limit is recovered by
taking $M_s\rightarrow \infty$.

The ${\cal N}=(1,1)$ little string theories with group $U(K)$, when
compactified on $T^3$, have a Coulomb branch which is $(\IR ^4\times
T^4)^{\otimes K}/S_K$ (more generally, $(\IR ^4\times T^4)^{{\rm
rank}(G)}/{\rm Weyl}(G)$), which probes the duality between type II
strings on $T^3$ and M theory on $T^4$.  The Coulomb branch $T^4$ has
metric $G^{AB}$ which is given exactly in terms of the $T^3$
compactification date by \metricf, with volume $M_s^2$.  There is a
similar statement for the ${\cal N}=(2,0)$ little string theory on
$T^3$, differing from the ${\cal N}=(1,1)$ case by a T-duality in one
of the $T^3$ cycles; the fixed volume $T^4$ in this context was also
discussed in \refs{\CGK, \CGKM}.  

The ${\cal N}=(1,0)$ little string theories associated with $K$ type
II or heterotic 5-branes at an $X_G\equiv \IC ^2/\Gamma _G$
singularity, when compactified on $T^3$, similarly probe M theory
geometry.  In the heterotic (or type II) cases, the M theory dual is
given by $K$ M2 branes with a transverse space $X_G\times K3$ (or
$X_G\times T^4$).  In both cases, M theory with a $X_G$ singularity
has an enhanced $G$ gauge symmetry and M2 branes, when sitting
directly on top of the $G$ singularity of $X_G$, can be interpreted as
small $G$ instantons.  In the heterotic (or type II) cases, these $K$
$G$-instantons have the fixed volume K3 (or $T^4$) as their four
spatial coordinates.  There is a moduli space for these instantons
given by their positions in these four spatial coordinates, as well as
their moduli for fattening up and rotating in $G$.

Thus, by the probe argument, the little string theory associated with
$K$ heterotic (or type II) branes at a $\IC ^2/\Gamma _G$ singularity,
when compactified on a $T^3$, has a compact Coulomb branch moduli
space of vacua which is exactly given by the moduli space of $K$
$G$-instantons on a K3 (or $T^4$) of volume $M_s^2$.  A quick check is
that the dimension of the Coulomb branch of the $T^3$ compactified
little string theories indeed agrees with the dimension of the moduli
space of $K$ $G$-instantons\foot{For a general 4
manifold with Euler character $\chi$ and signature $\sigma$, 
the dimension is $4KC_2(G)-\half|G|(\chi +\sigma )$. For $T^4$, $\chi
=\sigma =0$ and, for K3, $\chi =24$ and $\sigma = -16$.}
on $T^4$ or K3:  the type II cases indeed have $4(r_V+n_T)=4KC_2(G)$
and the heterotic cases indeed have $4(r_V+n_T)=4(KC_2(G)-|G|)$.  This
latter fact also played a role in the mirror symmetry of \EPGR.

Another check is to consider 
the limit $M_s^2\rightarrow
\infty$, where $T^4\rightarrow \IR ^4$ or $K3$ becomes a non-compact
piece of K3, and where the compactified little string theory goes over
to its 3d field theory limit.  In the type II cases, the resulting 3d
field theory has the quiver gauge group $\prod _{\mu =0}^r U(Kn_\mu)$,
based on the extended $G$-Dynkin diagram, which was indeed argued in
\refs{\ismir,
\berk} to have a quantum Coulomb branch which is the moduli space of
$K$ $G$-instantons on $\IR ^4$.  This theory was argued in \ismir\ to
have a hidden, global $G$ symmetry.  Because $M$ theory has $G$ gauge
symmetry even for finite $M_s$, the full compactified little string
theory is expected to also have this hidden global symmetry.  Similar
statements should hold in the heterotic cases.

Moduli spaces of instantons on $T^4$ or $K3$ have made a variety of
appearances in physics and mathematics, though usually with $G=U(N)$
as the gauge group.  In that case, the moduli space also depend on
$v_a=\int _{\Sigma _a} \Tr F$, where $\Sigma _a$ is a basis for the
two cycles of $T^4$ or K3.  In the present case, $G$ is a simple
$A,D,E$ group so $\Tr F=0$.  ($B$ fields can possibly still contribute
to $v_a\neq 0$, e.g. as in \obraneii.)

The moduli spaces of instantons obtained above have many interesting
singularities.  At these Coulomb branch singularities, there is an
attached Higgs branch, with an interacting 3d IR CFT at the intersection.

All of the above compact Coulomb branches are hyper-Kahler, with
$c_1=0$, and the sigma model metric is the unique one which is 
Ricci-flat.

\newsec{$T^2$ compactification: probing $F$ theory}

Compactifying the heterotic (or type II) little string theories to 4d
on a $T^2$ leads to quantum Coulomb branches which globally probe $F$
theory compactifications on a fixed volume, elliptically fibered K3
(or $T^4$).  For example, consider the $K=1$ case of
the heterotic little string theory (o), whose low-energy field theory
content is that of the world-volume of D3 branes in type I' on $T^2$.
This latter theory has a non-compact, quantum Coulomb branch which was
argued \refs{\sen,
\BDS}\ to locally probe the duality to F-theory on an 
elliptically fibered K3. The Coulomb branch in the 4d field theory is 
the non-compact complex $u$ plane, over which the photon coupling 
$\tau _{eff}(u)$ is fibered according to the Seiberg-Witten curve 
\SWii; the total space of $u$-plane base and $\tau _{eff}(u)$ fiber 
is a local, non-compact piece of K3.  This is the $M_s\rightarrow
\infty$ limit of the $T^2$ compactified little string theory.

Considering now the $T^2$ compactified little string theory for finite
$M_s^2$, the $u$-plane base is a compact box of volume $M_s^2$ (this
is the correct mass dimension for 4d scalars). As in sect. 3, reducing
a 6d $U(1)$ gauge field on a $T^2$ with metric $h_{ab}d\rho ^a d\rho
^b$ leads to scalars living on a dual $T^2$, with metric
$g_6^{-2}\sqrt{\det h}(h^{-1})^{ab}d\phi _a d\phi _b$, which has
volume $g_6^{-2}=M_s^2$ for all $h_{ab}$.  For $SU(2)$ rather than
$U(1)$, we mod out by the Weyl group $\phi _a\sim -\phi _a$, yielding
a 2d box of volume $M_s^2$. Considering the elliptic fiber $\tau (u)$
over the compact base as a dimensionless coordinate, the total space
of base and fiber is an elliptically fibered K3 of volume $M_s^2$.
This elliptically fibered K3 of fixed volume is that of the F-theory
dual to the 10d heterotic string on $T^2$.  As was the case there, the
parameter space \parmg\ of data in the $T^2$ compactification of the
heterotic string matches that of the fixed volume, elliptically
fibered K3s.  This can be regarded as a special case of the $T^3$
compactification considered in the previous sections, where one of the
radii is taken to infinity.  It is thus good that we again get a K3 of
volume $M_s^2$, since that was the case in the previous sections for
all radii.

More generally, $T^2$ compactifying the little string theories
associated with $K$ type II 5-branes at a $\IC ^2/\Gamma _G$
singularity leads to a compact Coulomb branch which is a
$2(r_V+n_T)=2KC_2(G)$ dimensional torus of unit volume in units of
$M_s$.  Including the $KC_2(G)$ complex dimensional elliptic fiber,
associated with the kinetic terms of the $KC_2(G)$ photons, the total
space is the moduli space of $K$ $G$-instantons on a $T^4$ of volume
$M_s^2$, where both the $T^4$ and the resulting instanton moduli space are
regarded as an elliptic fibration.

Compactifying on a $T^2$ the little string theories associated with
$K$ heterotic 5-branes at a $\IC ^2/\Gamma _G$ singularity leads to a
compact Coulomb branch which is a $2(r_V+n_T)=2(KC_2(G)-|G|)$
dimensional box, of unit volume in units of $M_s$.  Including the
fiber associated with the photons, the total space is an elliptically
fibered space which is exactly the moduli space of $K$ $G$-instantons
on an elliptically fibered K3 of volume $M_s^2$.

\bigskip
\centerline{{\bf Acknowledgments}}

I would like to thank M. Douglas, D. Morrison, R. Plesser, and
N. Seiberg for discussions.  This work was supported by UCSD grant
DOE-FG03-97ER40546 and IAS grant NSF PHY-9513835.

\listrefs

\end